\documentclass[a4paper,11pt]{article}
\usepackage{pos,latexsym,amsmath,amsfonts,amsthm,bm,axodraw2,tcolorbox,colortbl,multirow,hyperref,acronym,placeins}

\newcommand{\bq}{\begin{eqnarray}}
\newcommand{\eq}{\end{eqnarray}}

\newcommand{\Eulerconstant}{\gamma_{\mathrm{E}}}

\newcommand{\Dint}{D_{\mathrm{int}}}

\newcommand{\eps}{\varepsilon}

\newcommand{\loopnumber}{l}

\newcommand{\NE}{N_E}

\newcommand{\NV}{n}

\newcommand{\divisor}{p}

\newcommand{\Baikovvariable}{\sigma}

\newcommand{\preabs}{C_{\mathrm{abs}}}

\newcommand{\prebaikov}{C_{\mathrm{Baikov}}}

\newcommand{\differentialform}{\Psi}

\theoremstyle{plain}

\newboolean{longversion}
\setboolean{longversion}{false}

\title{An algorithm towards $\varepsilon$-factorising Feynman Integrals}

\author[a]{The $\varepsilon$-collaboration: Iris Bree}
\author[b]{Federico Gasparotto}
\author[a]{Antonela Matija\v{s}i\'c}
\author[a]{Pouria Mazloumi}
\author[a]{Dmytro Melnichenko}
\author[c]{Sebastian P\"ogel}
\author[a]{Toni Teschke}
\author*[d]{Xing Wang}
\author[a]{Stefan Weinzierl}
\author[e]{Konglong Wu}
\author[f]{Xiaofeng Xu}

\affiliation[a]{PRISMA Cluster of Excellence, Institut f{\"u}r Physik, Staudinger Weg 7,\\
Johannes Gutenberg-Universit{\"a}t Mainz, D - 55099 Mainz, Germany}

\affiliation[b]{Bethe Center for Theoretical Physics, Universität Bonn, D-53115 Bonn, Germany}

\affiliation[c]{Paul Scherrer Institut, CH-5232 Villigen PSI, Switzerland}

\affiliation[d]{School of Science and Engineering, The Chinese University of Hong Kong, Shenzhen, \\
      Longgang, 518172 Shenzhen, China}

\affiliation[e]{School of Physics and Technology, Wuhan University, \\
     No.299 Bayi Road, Wuhan 430072, China}

\affiliation[f]{Department of Physics, Xiamen University, \\
     Xiamen, 361005, China}     
      
\emailAdd{wangxing@cuhk.edu.cn}

\abstract{In this talk, we use several examples to elaborate on how a recently proposed algorithm~\cite{e-collaboration:2025frv,Bree:2025tug} can turn non-trivial Feynman integrals into an $\eps$-factorised manner, regardless of their hidden geometric essence. In particular, some extra details about three-loop banana integrals with unequal-mass configuration are provided.}

\FullConference{17th International Symposium on Radiative Corrections: Applications of Quantum Field Theory to Phenomenology (RADCOR2025)\\
5-10 October 2025\\
Puri, India\\}

\begin{document}
\maketitle

\section{Introduction}
Perturbative quantum field theory is essential in the precision era, not only in high-energy physics but also in gravitational-wave physics. As experiments become more and more precise, theoretical predictions using perturbative quantum field theory, where calculations of Feynman integrals are one of the bottlenecks, are challenged.  There are a lot of advances in this topic recently \cite{Chen:2020uyk,Chen:2022lzr,DHoker:2023khh,Marzucca:2023gto,delaCruz:2024xit,Baune:2024biq,Baune:2024ber,Jockers:2024uan,Gehrmann:2024tds,Pogel:2024sdi,Duhr:2024xsy,Gasparotto:2024bku,DHoker:2025szl,DHoker:2025dhv,Duhr:2025ppd,Duhr:2025tdf,Chaubey:2025adn,Bargiela:2025vwl,Carrolo:2026qpu,Duhr:2026ell,delaCruz:2026mas,Almeida:2026abr,Yang:2025ofz,delaCruz:2025szs,Jones:2025jzc,Chestnov:2025whi,Brammer:2025rqo,Correia:2025wtb,Liu:2022chg,Liu:2017jxz,Henn:2025xrc}. There are two major steps in modern techiniques of calculating Feynman integrals. The first one is to reduce Feynman integrals to a finite set of so-called master integrals based on integration-by-parts (IBP) identitities~\cite{Tkachov:1981wb,Chetyrkin:1981qh}, and the Laporta algorithm \cite{Laporta:2000dsw}. There are several public packages to implement IBP reductions \cite{vonManteuffel:2012np,Smirnov:2014hma,Maierhoefer:2017hyi,Wu:2023upw,Guan:2024byi,Lee:2013mka}. The second one is to solve the differential equation system of master integrals with respect to kinematics~\cite{Kotikov:1990kg,Kotikov:1991pm,Remiddi:1997ny,Gehrmann:1999as,Henn:2013pwa}. 

Both numeric and analytic calculations of Feynman integrals benefit dramatically if the dependence of $\eps$, the dimensional regulator, in the differential equations is entirely factorised out. Such a basis of master integrals and its differential equations is said to be in an $\eps$-factorised form~\cite{Henn:2013pwa}. Then one solves the $\eps$-factorised differential equations order by order in $\eps$ in terms of iterated integrals \cite{Chen} provided with suitable boundary values, which are simpler than the original integrals. 

It turns out that finding an $\eps$-factorised basis of master integrals has a relation to the associated geometry with a given family of Feynman integrals. There are many state-of-the-art calculations with non-trivial geometries, e.g., \cite{Henn:2025xrc,Adams:2018yfj,Honemann:2018mrb,Bogner:2019lfa,Muller:2022gec,Giroux:2022wav,Dlapa:2022wdu,Gorges:2023zgv,Delto:2023kqv,Jiang:2023jmk, Ahmed:2024tsg,Giroux:2024yxu,Duhr:2024bzt,Schwanemann:2024kbg,Marzucca:2025eak,Becchetti:2025oyb,Becchetti:2025rrz, Ahmed:2025osb, Chen:2025hzq,Coro:2025vgn,Pogel:2022vat,Pogel:2022yat,Pogel:2022ken,Duhr:2022dxb,Forner:2024ojj,Frellesvig:2024rea,Duhr:2025lbz,Maggio:2025jel,Duhr:2025kkq,Pogel:2025bca,Duhr:2024uid,Wang:2024ilc}. Recently, an algarithm was proposed in \cite{e-collaboration:2025frv,Bree:2025tug}, inspired by Hodge theory, to decompose finding an $\eps$-factorised basis into two steps. This method is applicable in a unified way regardless of the associated geometries and, hence, opens a new window to calculate Feynman integrals systematically and efficiently. And it also sheds light on the deep connection between perturbative quantum field theory and mathematics, in particular, algebraic geometry. 

In this talk, we use two non-trivial examples to illustrate the basic ideas of the two steps of this algorithm. Some extra information compared to \cite{Bree:2025tug,Pogel:2025bca} are provided.

\section{A Sketch of the Method}
We are interested in the family of Feynman integrals
\bq
\label{def_feynman_integral}
 I_{\nu_1 \dots \nu_{\NE}}\left(D, x \right)
 & = &
 e^{\loopnumber \eps \Eulerconstant} (\mu^2)^{\nu-\frac{\loopnumber D}{2}}
 \int \prod\limits_{r=1}^{\loopnumber} \frac{{\rm d}^Dk_r}{i \pi^{\frac{D}{2}}} 
 \prod\limits_{j=1}^{\NE} \Baikovvariable_j^{-\nu_j},
\eq
where $\eps=(\Dint-D)/2$ (with $\Dint \in {\mathbb Z}$) 
denotes the dimensional regularisation regulator,
$\gamma_E$ denotes the Euler-Mascheroni constant, 
$\mu$ is an arbitrary scale introduced to render the Feynman integral dimensionless. The integral family depends on dimensionless kinematic variables $x$. Given the inverse propagators $\sigma_i$'s, there are infinite integrals in this integral family due to arbitrary indexes, $\nu_i$. With the IBP relations, master integrals in a given sector form a finite vector space, denoted as $V^n$, where $n$ will be manifest later on.

The recently proposed algorithm\cite{e-collaboration:2025frv,Bree:2025tug} towards an $\eps$-factorised basis is made of two steps. Firstly, given a possible starting basis (in the interested top sector) $I$, we build a new basis $J$ determined by the filtration criteria. It is related to the starting basis by a rotation:
\begin{equation}
	J=R_1^{-1} I\,,
\end{equation}
such that 
\begin{equation}
	{\rm d} J(\varepsilon, x)=\sum_{k=-n}^1 \varepsilon^k \hat{A}^{(k)}(x)J(\varepsilon, x)\,.
\end{equation}
We say the differential equation in the basis $J$ is in Laurent polynomial form. To obtain such a basis, we adopt the ``separate and conquer'' strategy to start with the maximal cut, and we resort to studying the structures of the master integrands instead of integrals under the maximal cut. The maximal cut of a Feynman integral is most easily analysed in the loop-by-loop Baikov representation \cite{Frellesvig:2017aai}:
\bq\label{eq:Bairep}
 \operatorname{Res}\limits_{{\mathcal C}_{\mathrm{maxcut}}} I_{\nu_1 \dots \nu_{\NE}}\left(D, x \right)
 & \sim & 
 \int {\rm d}^{\NV}z \;
 \prod\limits_{i \in I_{\mathrm{all}}} \left[ \divisor_i\left(z\right) \right]^{\alpha_i}.
\eq
Here, $n$ is the number of remaining Baikov variables. The integrand in the above contains the so-called twist function $u(z)$ in the intersection theory~\cite{Mastrolia:2018uzb,Frellesvig:2019uqt}. The central object in \cite{e-collaboration:2025frv, Bree:2025tug} is the master integrands after homogenisation:
\begin{equation}\label{eq:masint}
	\Psi_{\mu_0 \ldots \mu_{N_D}}[Q] = {\color[RGB]{26,22,197}C_{\text {Baikov }} }\cdot {\color[RGB]{26,22,197} C_{\text {abs }}}\cdot {\color[RGB]{144,0,33} C_{\text {rel }}(\mu)} \cdot  {\color[RGB]{144,0,33} C_{\text {clutch }}(\mu) }\cdot  U(z) \cdot \hat{\Phi}_{\color[RGB]{144,0,33}\mu_0 \ldots \mu_{N_D}}[Q]\cdot \eta.
\end{equation}
Here, $N_D=|I_{\mathrm{all}}|$ is the number of all polynomials in the twist, and $U(z)$ is the homogeneous version of $u(z)$, and $\eta$ is the homogenised integration measure in \eqref{eq:Bairep}. The concept of filtrations introduces refined order numbers on top of the Laporta algorithm. The upshot is that master integrands in a given sector, forming a finite-dimensional vector space, denoted as $H_\omega^n$, have multi-layered structures. Complexity in different layers is ordered according to the filtration index. 
\begin{enumerate}
	\item The prefactor $C_{\rm Baikov}$ from the initial Baikov representation of a master integral tells us what $C_{\rm abs}$ should be such that ${\color[RGB]{26,22,197}C_{\rm Baikov}\cdot C_{\rm abs}}$ is pure. $C_{\rm abs}$ is rational in $\eps$ and may involve a simple power-like function in kinematics, but not any Baikov variables. 
	\item Then, we exhaust the candidate integrands according to the (top-down) weight filtration, $w=n+r$, with $r$ being the number of non-zero residues that the candidate can take. The ${\color[RGB]{144,0,33} \mu=(\mu_0, \cdots,\mu_{N_D}) }$ index determines the other two prefactors, ${\color[RGB]{144,0,33} C_{\text {rel }}(\mu)} $ and ${\color[RGB]{144,0,33} C_{\text {clutch }}(\mu) }$, where we do not repeat their definitions for brevity. Candidates, according to the weight filtration, may have linear relations among them, so one should run IBP reduction at this layer. 
	\item In the layer specified by the weight $w$, a further (left-right) filtration, $p=w-o$, where $o$ is the pole order of a candidate integrand describing how singular it is, is used to order those candidates. One may also need to run IBP at this specific sub-space. 
\end{enumerate}

After we exhaust those finite master integrands by these two filtrations together with IBP relations, we obtain the desired candidate master integrands. Translating them back to the Feynman side leads us to the basis $J$. There are several points worth emphasising: 1) the matrix elements in $R_1$, and the new connection matrices $\hat{A}^{(k)}(y)$'s are always rational; 2)  the new connection matrices $\hat{A}^{(k)}(y)$'s have a block-triangular pattern organised by the filtration criteria (or simply complexity). As a result, the outcome at this step is already very useful in practice; 3) sometimes, the ``unwanted'' terms $\hat{A}^{(-n)}, \cdots, \hat{A}^{(0)}$ vanish after this step. Namely, the obtained basis after the first step is already $\eps$-factorised, e.g., the first example. This is very common in Feynman integrals of multiple-polylogarithmic nature.

The second step is to introduce a further algorithmic rotation to get rid of the ``unwanted'' terms $\hat{A}^{(-n)}, \cdots, \hat{A}^{(0)}$ after the first step if necessary:
\begin{equation}\label{eq:basisK}
	K=R_2^{-1} J=R_2^{-1} R_1^{-1} I\,.
\end{equation}
Since the $\eps$ dependence is in the Laurent way and the structures of $\hat{A}^{(-i)}$ for $0\leq i\leq n$ are block-triangular, the requirement of getting rid of them is decomposed into step-by-step constraints (simpler than differential equations of the whole system):
\begin{equation}\label{eq:R2decom}
	R_2=R_2^{(-n)} R_2^{(-n+1)} \ldots R_2^{(-1)} R_2^{(0)}\,, 
\end{equation}
where all matrices $R_2^{(k)}$'s are lower block-triangular. We first do the rotation $R_2^{(-n)}$ (if necessary), followed by $R_2^{(-n+1)}$, so on and so forth, due to the inverse in \eqref{eq:basisK}. The superscipts in \eqref{eq:R2decom} are called the $B$-order. Roughly speaking, it corresponds to the Laurent expansion order of $\eps$. However, strictly speaking, lower-order ones can contribute to higher ones when combined with $\hat{A}^{(1)}$. For the rigorous definition, please refer to \cite{e-collaboration:2025frv,Bree:2025tug}. We will use examples to illustrate instead. There are several points worth emphasising about this step:
\begin{itemize}
	\item The rotation matrix $R_2^{(-n)}$ is related to periods of the relevant geometries of the Feynman integrals. Nevertheless, we do not have to know the geometric information explicitly. The algorithm applies to all Feynman integrals, as far as we know, in a unified way. 
	\item Solutions of constraints, and hence the matrix elements in the rotations, are transcendental. We do not have to solve the constraints analytically. Instead, one can solve them numerically, which is simpler than solving the Feynman integrals numerically directly. 
	\item Knowing analytic solutions of the rotation matrices can deepen our understanding of the mathematical structures of given Feynman integrals. 
\end{itemize}

\section{Appetiser Example: Pentabox}
\begin{figure}
	\centering
	\includegraphics[width=0.3\textwidth]{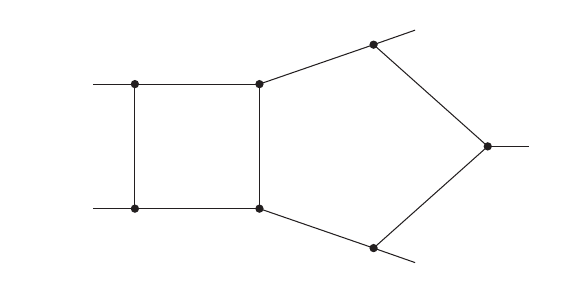}
	\caption{\label{fig:db}Massless onshell pentabox.}
\end{figure}

The appetiser example is the massless on-shell pentabox integral family~\cite{Gehrmann:2018yef}, shown in Fig.~\ref{fig:db}. 
This is an example with three master integrals in the top sector.
The inverse propagators are:
\begin{align}
 \sigma_1 & = -k_1^2 ,
 &
 \sigma_2 & = -\left(k_{1}+p_1\right)^2 ,
 &
 \sigma_3 & = -\left(k_{1}+p_{12}\right)^2,
 \nonumber \\
 \sigma_4 & = -\left(k_{1}+p_{123}\right)^2,
 &
 \sigma_5 & = - k_2^2,
 &
 \sigma_6 & = -\left(k_{2}+p_{123}\right)^2,
 \nonumber \\
 \sigma_7 & = -\left(k_{2}+p_{1234}\right)^2,
 &
 \sigma_8 & = -\left(k_1-k_2\right)^2,
 &
 \sigma_9 & = -\left(k_{1}+p_{1234}\right)^2,
 \nonumber \\
 \sigma_{10} & = -\left(k_{2}+p_{1}\right)^2,
 &
 \sigma_{11} & = -\left(k_{2}+p_{12}\right)^2,
\end{align}
with $p_{iI} = p_i + p_I$ iteratively.  We set $\mu^2 = s_{45}$ to render the integral family dimensionless, and it depends on the following four dimensionless kinematic variables:
\bq
	x_1 = \dfrac{s_{12}}{s_{45}}, \quad x_2 = \dfrac{s_{23}}{s_{45}}, \quad x_3 = \dfrac{s_{34}}{s_{45}}, \quad x_4 = \dfrac{s_{15}}{s_{45}}, 
\eq
where $s_{ij}=2 p_i \cdot p_j$. A loop-by-loop Baikov representation on the maximal cut is
\begin{align}\label{eq:BKpd}
 \operatorname{Res}_{\mathcal{C}_{\rm maxcut}}\,I_{11111111000} =&\, \prebaikov \int \frac{{\rm d}z_1}{2\pi i}
 z_1^{-1-\eps}\left(z_1-1\right)^{\eps}\\
\times &  \left(\left(1-x_1-x_2\right) z_1^2+\left(x_3 - x_2 x_3+x_1\left(x_2 -x_4 \right) + x_4\right) z_1+x_3 x_4 \right)^{-1-\eps},\nonumber
\end{align}
where $z_1=\sigma_9/s_{45}$ and 
\begin{align}
\prebaikov &= \dfrac{32 i\,e^{2 \eps \Eulerconstant}  \pi^{5}}{\Gamma\left( -2 \eps\right)} (x_1x_2)^{-1-\eps}  \Delta^{\frac{1}{2}+\eps}\,.
\end{align}
See \cite{Bree:2025tug} for the expression of $\Delta$. By setting\footnote{This is a bit different from that in \cite{Bree:2025tug}.} $\preabs= \eps^4$, the product $\prebaikov \cdot \preabs$ is pure of transcendental weight zero. The minimal twist function in the projective space reads from \eqref{eq:BKpd} as 
\begin{align}
    U(z_0, z_1) &= P_0^{2\eps} P_1^{-\eps} P_2^{\eps} P_3^{-\eps}\,,\qquad \text{with}\qquad P_0=z_0, \quad P_1=z_1, \quad P_2=z_1 - z_0\,, \nonumber \\
	P_3&=\left(1-x_1-x_2\right) z_1^2+\left(x_3 - x_2 x_3+x_1\left(x_2 -x_4 \right) + x_4\right) z_0 z_1+x_3 x_4 z_0^2\,. 
\end{align}
By solving ${\rm d} \ln U = 0$, we find that $\dim H^1_\omega = 3 = \dim V^1\,$. 
So we do not need to consider symmetry loss from integrands to integrals or super-sectors. The projective measure $\eta$ in this example is $\eta=z_0 {\rm d} z_1 - z_1 {\rm d} z_0 $ of homogeneous degree 2, while the projective minimal twist is of homogeneous degree 0. So, we are looking for  rational functions $\hat{\Phi}_{\mu_0 \mu_1 \mu_2 \mu_3}[Q]$  with homogeneous degree $(-2)$. We start from the top layer of the vector space of differential forms. There are, in total, 5 zeros determined by $P_0 - P_3$, which indicate five possibilities to localise on a point. After that, one can not take one more residue any longer. Running the Laporta algorithm on this layer will reduce them to three master integrands. A possible choice for the master integrands is
\begin{equation}\label{master_integrands_pentabox}
	\begin{aligned}
 \differentialform_{0001}\left[1 \right]
  & =   C_{\rm Baikov}\cdot C_{\rm abs} \cdot {\color[RGB]{144,0,33}(-1)} \cdot U \cdot\frac{1}{P_3}\cdot \eta \,\,\quad \longleftrightarrow \quad   J_1 = - \eps^4 I_{11111111(-1)00}\,,
  \\
 \differentialform_{0101}\left[z_0 \right]
 & = 
 C_{\rm Baikov}\cdot C_{\rm abs} \cdot {\color[RGB]{144,0,33} 1 }\cdot  U \cdot \frac{z_0}{P_1P_3}\cdot \eta \;\;\quad \longleftrightarrow \quad  J_2 = \eps^4 I_{11111111000}\,,
  \\
 \differentialform_{1001}\left[z_1 \right]
 & =  C_{\rm Baikov}\cdot C_{\rm abs} \cdot {\color[RGB]{144,0,33}(-2)} \cdot  U  \frac{z_1}{P_0P_3}\cdot \eta \quad \longleftrightarrow \quad J_3 = - 2\eps^4 I_{11111111(-2)00}\,,	
	\end{aligned}
\end{equation}
where the colored prefactors in the middle parts are determined by the index of rational parts, which we bypass for brevity. The rightmost parts are their Feynman side correspondence. There is no need to go one layer down, since we already have three linearly independent candidates. If one insists on going one layer down, then the same results will be obtained. It turns out that the above three master integrals \textit{almost} enjoy the desired $\eps$-factorised differential equations with respect to all kinematic variables. A trivial step 2 rotation can be found instantly:
\begin{equation}
	K_1 = J_1\,,\qquad K_2 = J_2\,,\qquad K_3 = x_1 x_2 \cdot J_3\,.
\end{equation}

\section{Three-loop Banana with Unequal masses}
\begin{figure}
\begin{center}
\includegraphics[scale=0.75]{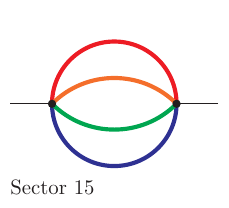}
\includegraphics[scale=0.75]{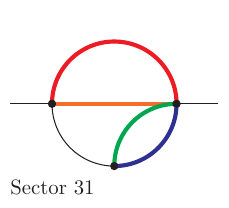}
\includegraphics[scale=0.75]{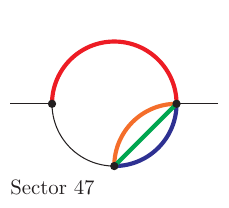}
\includegraphics[scale=0.75]{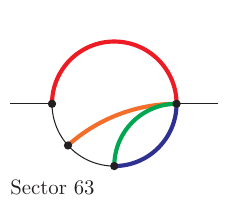}
\end{center}
\caption{
The banana integral (the leftmost sector 15) together with its relevant super-sectors.
}
\label{fig:super_sectors}
\end{figure}

Now, we turn to the main-course example of applying the method: the three-loop banana integral with four distinct masses shown in Fig.~\ref{fig:super_sectors}, which is highly nontrivial. The inverse propagators are given by
\begin{equation}
	\begin{aligned}
 \sigma_1 & = -k_1^2 + m_1^2,
 &
 \sigma_2 & = -k_2^2 + m_2^2,
 &
 \sigma_3 & = -k_3^2 + m_3^2,
 \nonumber \\
 \sigma_4 & = -\left(k_1+k_2+k_3-p\right)^2 + m_4^2,
 &
 \sigma_5 & = -\left(k_1+k_2-p\right)^2,
 &
 \sigma_6 & = -\left(k_1-p\right)^2,
 \nonumber \\
 \sigma_7 & = -\left(k_1+k_2\right)^2,
 & 
 \sigma_8 & = -\left(k_1+k_3\right)^2,
 &
 \sigma_9 & = -\left(k_2+k_3\right)^2.
\end{aligned}  
\end{equation}
The three-loop banana integral has sector id $15$, corresponding to the propagator set $S=\{1,2,3,4\}$. We set $\mu^2 = -s = -p^2$ as normalisation of the integral family, such that this family depends on the following four dimensionless variables:
\begin{equation}
	y_1 = \frac{m_1^2}{-s}\,,\quad y_2 = \frac{m_2^2}{-s}\,,\quad y_3 = \frac{m_3^2}{-s}\,,\quad y_4 = \frac{m_4^2}{-s}\,.
\end{equation} 

There are, in total, 15 master integrals relevant to sector 15, 4 of which are tadpoles. One can start with the loop-by-loop Baikov representation of $I_{1111\nu_5 \nu_6 000}$ under the banana cut, to derive the minimal twist function $u(z_1, z_2)$, 
\begin{equation}
	\operatorname{Res}_{\mathcal{C}_{\rm maxcut}}\,I_{1111\nu_5\nu_6000} = \left(-s\right)^{1+4 \varepsilon}\underbrace{\left(2\pi\,e^{ \gamma_E \varepsilon}\frac{\Gamma(-\eps) }{\Gamma(-2\eps)}\right)^3}_{C_{\rm Baikov}} \int \frac{{\rm d} z_1  {\rm d} z_2}{(2 \pi i)^2} u(z_1, z_2) z_1^{-\nu_5} z_2^{-\nu_6}
\end{equation}
with $z_1 = \frac{\sigma_5}{-s}$ and $z_2 = \frac{\sigma_6}{-s}$. The homogenised version of the twist function, $U(z_0, z_1, z_2)$ reads 
\bq
 U
 & = &
 P_0^{4\eps}
 P_1^{\eps}
 P_2^{\eps}
 P_3^{-\frac{1}{2}-\eps}
 P_4^{-\frac{1}{2}-\eps}
 P_5^{-\frac{1}{2}-\eps}.
\eq
This twist function can be generalised to $l$ loops as follows with $y_i = m_i^2/(-s)$:
\bq
 U(z_0, \cdots, z_{l-1})
 & = &
 P_0^{(l+1)\eps}
 P_1^{\eps}\,
 P_{l}^{-\frac{1}{2}-\eps}\,
\left(\,\prod_{i= 2}^{l-1}  P_{i}^{\eps}\, P_{l-1+i}^{-\frac{1}{2}-\eps}\,\right)\,
 P_{2l-1}^{-\frac{1}{2}-\eps}
 ,
\eq
with
\begin{equation}
	\begin{aligned}
		P_i &= z_i\,,\qquad i \in \{ 0, 1, \cdots, l-1 \}\,,\\
		P_l &= z_1^2 + 2(y_l+y_{l+1})z_1 + (y_l - y_{l+1})^2\,,\\
		P_{l-1+i} &= (z_{i+1} - z_i)^2 + 2 y_{l+1-i}(z_{i+1} + z_i) + y_{l+1-i}^2\,, \qquad i \in \{ 2, 3 \cdots, l-1 \}\,,\\
		P_{2l-1} &= z_{l-1}^2 - 2(1-y_1)z_{l-1} + (1+y_1)^2\,.
	\end{aligned}
\end{equation}

Back to the three-loop scenario, the twist function tells us that
\begin{equation}
	\dim H_\omega^2 = 13 \quad > \quad  11 = \dim V^2.
\end{equation}
This is because the twisted side also takes super sectors into account. 
The super-sectors 31 and 47 have one master integral each. A possible basis for these super-sectors on the Feynman integral side is $\{I_{111110000}\,, I_{111101000}\,\}$. The super-sector 63 is reducible and does not have any master integral. These two additional master integrals make up for the difference in dimensions of $V^2$ and $H_\omega^2$. 

\subsection{Basis $J$}
\begin{figure}[!htp]
\begin{center}
\includegraphics[scale=0.4]{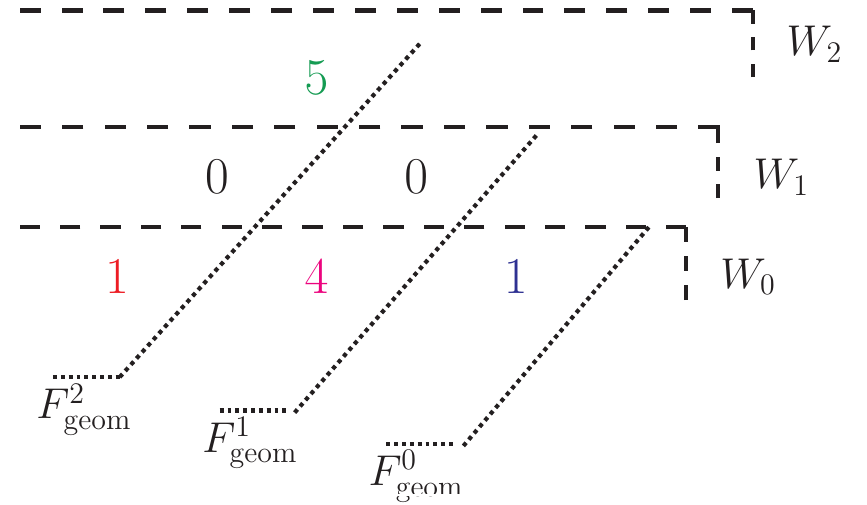}
\end{center}
\caption{
The dimensions of decomposed subspaces within the top sector. The subscript ``geom'' means that the geometric filtration is used here.
}
\label{fig:bananahodge}
\end{figure}

Although it is a non-trivial example to illustrate how to perform the first step in the algorithm, \cite{Pogel:2025bca} explains this part in detail, and we suppress this step in the proceeding due to constraints on pages, while focusing on elaborating a bit on the second step. After the first step, we obtain the following basis $J$, including the tadpoles.
\begin{equation}\label{eq:bananaJ}
\begin{aligned}
 J_1
  & =  
 \eps^3 I_{011100000},\,\,\,J_2
   =   
 \eps^3 I_{101100000},\,\,\,
 J_3
  =  
 \eps^3 I_{110100000},\,\,\,
 J_4
  =  
 \eps^3 I_{111000000}, \\
 {\color[RGB]{191,58,47}J_5 }
 & {\color[RGB]{191,58,47} =  
 \eps^3 \; I_{111100000}, }
 \\
 {\color[RGB]{192,51,111}
 J_{5+i} }
 & {\color[RGB]{192,51,111}=  
 \frac{1}{\eps} y_i \frac{\partial}{\partial y_i} J_5, \quad i=1, 2, 3, 4\,,}
  \\
 {\color[RGB]{45,47,117}
 J_{10} }
 & {\color[RGB]{45,47,117}=  
 \left.
 \eps^3 \left[ 6 I_{1111(-1)0000} - 8 I_{11110(-1)000} + 3 I_{111100000} \right] \right|_{\mathrm{sector} \; 15},}
  \\
 {\color[RGB]{45,47,117}
 J_{11} }
 & {\color[RGB]{45,47,117}=  
 \left.
 \eps^3 \left[3 I_{1111(-1)0000} - I_{11110(-1)000} + 3 y_2 I_{111100000}
 \right] \right|_{\mathrm{sector} \; 15},}
  \\
 {\color[RGB]{45,47,117}
 J_{12} }
 & {\color[RGB]{45,47,117}=  
 \eps^3 \left( 1+y_1 \right) 
 \left. \left( 3 I_{1111(-1)1000} + 3 y_2 I_{111101000} - I_{111100000} \right)\right|_{\mathrm{sector} \; 15},}
  \\
 {\color[RGB]{45,47,117} 
 J_{13} }
 & {\color[RGB]{45,47,117}=  
 \eps^3 \left( y_3-y_4 \right) \left. \left( I_{11111(-1)000} + y_2 I_{111110000} \right)\right|_{\mathrm{sector} \; 15}, }
 \\
 {\color[RGB]{45,47,117} 
 J_{14} }
 & {\color[RGB]{45,47,117} =  
 \eps^3 \left( 1+y_1 \right) \left( y_3-y_4 \right) \left. \left( y_2 I_{111111000} + I_{111110000} + I_{111101000} \right)\right|_{\mathrm{sector} \; 15},}
  \\
 {\color[RGB]{56,141,87}
 J_{15} }
 & {\color[RGB]{56,141,87}=  
 \frac{1}{16 \eps^2} \left[ y_1 \frac{\partial}{\partial y_1} + y_2 \frac{\partial}{\partial y_2} + y_3 \frac{\partial}{\partial y_3} + y_4 \frac{\partial}{\partial y_4} \right]^2 J_5. }
\end{aligned}
\end{equation}
The colours in the above correspond to those in Fig.~\ref{fig:bananahodge}.

%
%

\subsection{Constraints}
The basis $J$ is not yet $\eps$-factorised. Hence, we need step 2 in this example. This section is devoted to the constraints in the rotation matrix $R_2$ in step 2.  By $B$-ordering, see details in \cite{Bree:2025tug}, the matrix is decomposed as
\begin{equation}
	R_2 = R^{(-2)}_2 R^{(-1)}_2 R^{(0)}_2,\quad \text{and}\quad R^{(-2)}_2 R^{(-1)}_2 R^{(0)}_2\, K = J.
\end{equation}
Transformations in this step always maintain the $\eps$-factorised mixing with the tadpoles. Hence, from now on, we will simply suppress matrix elements related to tadpoles. However, we keep the master integral number index as \eqref{eq:bananaJ}. 

Below, we elaborate on the rotation $R_2^{(-2)}$, and comment on the other two at the end of this section. In \eqref{A_ldegree}, we show on the left panel the lowest non-vanishing order in $\eps$ for each entry of the connection matrix for the basis $J$ given in the above without tadpoles:
\bq
\label{A_ldegree}
 \left(
\begin{array}{r|rrrrrrrr|rr}
  - & 1 & 1 & 1 & 1 & - & - & - & - & - & - \\
 \hline
  -1 & 0 & 0 & 0 & 0 & 1 & 1 & 1 & 1 & 1 & 1 \\
  -1 & 0 & 0 & 0 & 0 & 1 & 1 & 1 & 1 & 1 & 1 \\
  -1 & 0 & 0 & 0 & 0 & 1 & 1 & 1 & 1 & 1 & 1 \\
  -1 & 0 & 0 & 0 & 0 & 1 & 1 & 1 & 1 & 1 & 1 \\
  0 & 1 & 1 & 1 & 1 & 1 & 1 & 1 & 1 & - & - \\
  0 & 1 & 1 & 1 & 1 & 1 & 1 & 1 & 1 & - & - \\
  0 & 1 & 1 & 1 & 1 & - & 1 & 1 & - & 1 & - \\
  0 & 1 & 1 & 1 & 1 & 1 & 1 & - & 1 & 1 & - \\
 \hline
  0 & 1 & 1 & 1 & 1 & - & - & 1 & 1 & 1 & - \\
  -2 & -1 & -1 & -1 & -1 & 0 & 0 & 0 & 0 & 0 & 0 \\
 \end{array}
 \right),\qquad \begin{aligned}
 &\mbox{$B$-order}\ (-2)
 :
 \scalebox{0.9}{$\left( \begin{array}{ccc}
 0 & - & - \\
 -1 & 0 & - \\
 -2 & -1 & 0 \\
 \end{array} \right)$}.
 & 
\end{aligned}
\eq
The right panel of \eqref{A_ldegree} shows the block structures of $B$-order $(-2)$ in the above connection matrix. The transformation $R_2^{(-2)}$ removes terms of $B$-order $(-2)$ in the above, whose block structures read
{\footnotesize
\bq
\label{Rm2_ldegree}
 \left(
 \begin{array}{r|rrrrrrrr|rr}
 R^{(-2)}_{55} & 0 & 0 & 0 & 0 & 0 & 0 & 0 & 0 & 0 & 0 \\
 \hline
 \frac{1}{\eps} R^{(-2)}_{65} & R^{(-2)}_{66} & R^{(-2)}_{67} & R^{(-2)}_{68} & R^{(-2)}_{69} & 0 & 0 & 0 & 0 & 0 & 0 \\
  \frac{1}{\eps} R^{(-2)}_{75} & R^{(-2)}_{76} & R^{(-2)}_{77} & R^{(-2)}_{78} & R^{(-2)}_{79} & 0 & 0 & 0 & 0 & 0 & 0 \\
 \frac{1}{\eps} R^{(-2)}_{85} & R^{(-2)}_{86} & R^{(-2)}_{87} & R^{(-2)}_{88} & R^{(-2)}_{89} & 0 & 0 & 0 & 0 & 0 & 0 \\
 \frac{1}{\eps} R^{(-2)}_{95} & R^{(-2)}_{96} & R^{(-2)}_{97} & R^{(-2)}_{98} & R^{(-2)}_{99} & 0 & 0 & 0 & 0 & 0 & 0 \\
  0 & 0 & 0 & 0 & 0 & 1 & 0 & 0 & 0 & 0 & 0 \\
 0 & 0 & 0 & 0 & 0 & 0 & 1 & 0 & 0 & 0 & 0 \\
 0 & 0 & 0 & 0 & 0 & 0 & 0 & 1 & 0 & 0 & 0 \\
  0 & 0 & 0 & 0 & 0 & 0 & 0 & 0 & 1 & 0 & 0 \\
 \hline
  0 & 0 & 0 & 0 & 0 & 0 & 0 & 0 & 0 & 1 & 0 \\
  \frac{1}{\eps^2} R^{(-2)}_{F5} & \frac{1}{\eps} R^{(-2)}_{F6} & \frac{1}{\eps} R^{(-2)}_{F7}  & \frac{1}{\eps} R^{(-2)}_{F8}  & \frac{1}{\eps} R^{(-2)}_{F9}  & 0 & 0 & 0 & 0 & R^{(-2)}_{FE} & R^{(-2)}_{FF} \\
 \end{array}
 \right).
\eq
}
Here, we used hexadecimal notation for the indices. All the functions $R_{ij}^{(-2)}(y)$'s in the ansatz are yet to be determined. We sketch them in the following.
\begin{enumerate}
	\item Implementing this rotation develops $\eps^0$ terms in position\footnote{The whole system is $15\times 15$. } $(5, 5)$. To remove $(\cdots){\rm d} y_i$ require that 
	\begin{equation}\label{eq:R65}
		R_{(i+1)5}^{(-2)} = y_i \partial_i \psi_0,\quad i = 1, 2, 3, 4\,,
	\end{equation}
	where $\partial_i = \partial/\partial y_i$ and we have renamed $R_{55}^{(-2)}$ as $\psi_0(y)$. 
	\item Plugging \eqref{eq:R65} into the differential equation system, to remove the $\eps^{-1}$ terms of the entries $(6,5),(7,5),(8,5),(9,5)$ results into second-order constraints on $\psi_0$ with a new unknown function $R_{F5}^{(-2)}$:
	\begin{equation}
		-\Big[ y_i \partial_i^2  + \partial_i + 1  \Big]\psi_0 = 2\sum_{j=1}^4 y_j\partial_j \psi_0+ 16R^{(-2)}_{F5}, \quad i = 1, 2, 3, 4\,.
	\end{equation} 
	There are also expressions for the mixed derivatives, which are more lengthy\footnote{These mixed-derivative constraints can be used as cross-checking conditions.}. We can rewrite the unknown function $R_{F5}^{(-2)}$ in terms of $\psi_0$ and its derivatives:
	\begin{equation}
	\label{eq:RF5}
	\begin{aligned}
		R^{(-2)}_{F5} &= -\frac{1}{64}\bigg[4 + \sum_{i=1}^4y_i \partial_i^2 + (8y_i+1)\partial_i  \bigg]\psi_0 \,. 
	\end{aligned}
\end{equation}
    \item To remove $\eps^{-2}$ terms in position $(15, 5)$ leads us to first-derivative constraints on $R^{(-2)}_{F5}$. Combining with \eqref{eq:RF5} gives us third-order constraints on $\psi_0$, which we suppress here for shorthand. As long as we know $\psi_0$, we know $R_{55}^{(-2)}, R_{65}^{(-2)}, R_{75}^{(-2)}, R_{85}^{(-2)}, R_{95}^{(-2)}$ and $R_{F5}^{(-2)}$. 
\end{enumerate}
In summary, to remove the first column in the $B$-order (-2) matrix is equivalent to finding $\psi_0$, which is the solution of a set of operators. Those third-order operators annihilating $\psi_0$ form the so-called Picard-Fuchs ideal. In the equal-mass limit, they degenerate to the familiar Picard-Fuchs operator relevant to the equal-mass three-loop banana integral. One can view the above procedure as a method to derive the Picard-Fuchs ideal, which annihilates the periods for geometry. 

One does not need to solve $\psi_0$ with generic $y_1, y_2, y_3$ and $y_4$. Instead, one can choose a specific parameterisation and use the Frobenius method to solve $\psi_0$ from the specifically degenerated Picard-Fuchs operator. Here, we can write down the ansatz for solutions of the Picard-Fuchs ideal, i.e., periods, inspired by \cite{Pogel:2022yat,Jockers:2024uan}, and \textit{check they satisfy all the constraints explicitly}. With the help of the shorthand notation:
\bq
\bm{n}\; = \;(n_1,n_2,n_3,n_4),  \qquad 
 \bm{y}^{\bm{n}}
 \; = \;
 y_1^{n_1} y_2^{n_2} y_3^{n_3} y_4^{n_4},
 \qquad 
 \left| \bm{n} \right| \; = \; n_1+n_2+n_3+n_4,
\eq
we can write down $\psi_0$ and the other four solutions of the Picard-Fuchs ideal in a compact way as:
\bq
\label{def_Frobenius}
 \psi_{0} = 
 \sum\limits_{n=0}^\infty 
 \sum\limits_{|\bm{n}|=n}
 a_{\bm{n}} \bm{y}^{\bm{n}}\,;\qquad 
 \psi_{1,j} = 
 \frac{1}{2\pi i}
 \sum\limits_{n=0}^\infty 
 \sum\limits_{|\bm{n}|=n}
 \left[
 a_{\bm{n},j} 
 +
 a_{\bm{n}} 
 \ln y_j
 \right]
 \bm{y}^{\bm{n}},
 \quad
 1 \; \le \; j \; \le \; 4\,.
\eq
The coefficients are given by
\bq
\label{Frobenius_coeffs}
 a_{\bm{n}}
  = 
 \left(-1\right)^{\left|\bm{n}\right|}
 \left( \frac{\left|\bm{n}\right|!}{n_1! n_2! n_3! n_4!} \right)^2,\quad a_{\bm{n},j} = 2 \left[ S_1\left(\left|\bm{n}\right|\right) - S_1\left(n_j\right) \right]\,a_{\bm{n}},
\eq
where $S_m(n)$ denotes the harmonic sum $S_m\left(n\right) =  \sum\limits_{j=1}^n j^{-m}.$ With the above explicit results, we have
\begin{equation}
	\label{eq:polar}
	R^{(-2)}_{(i+5)5} =  \sum_{n=0}^{\infty} \sum_{|{\bm n}|=n} n_i a_{\bm n} {\bm y}^{\bm n}\,,\quad i \in \{1, 2, 3, 4\}\,;\qquad\quad R^{(-2)}_{F5} = \frac{1}{16}\sum_{n=0}^{\infty} \sum_{|{\bm n}|=n} n^2 a_{\bm n} {\bm y}^{\bm n}\,.
\end{equation}

Now, we can turn to the other unknown functions in $R_2^{(-2)}$. There are two points to be emphasised: 1) constraints about the sixteen functions $R^{(-2)}_{66}, \dots, R^{(-2)}_{99}$ mix with each other, so we do not investigate them separately; 2) like the case of the first colume,  elements in the last row can be expressed with $R^{(-2)}_{66}, \dots, R^{(-2)}_{99}$ and their derivatives. It is natural to choose the variable $y_i$ in the constraints when we are talking about the $(i+5)$-th row, because the $(i+5)$-th master integral has a double propagator depending on $m_i^2$. In this context, we achieve
\begin{equation}\label{eq:RFj}
	R_{F j}^{(-2)} = -\frac{1}{64} \sum_{i=1}^4\left(\partial_i\, +\, \partial_i\,\log \psi_0 + 8 \right)R_{(5+i) j}^{(-2)}, \quad j \in\{6,7,8,9\} .
\end{equation}
One can follow \cite{Pogel:2025bca} to solve $R^{(-2)}_{66}, \dots, R^{(-2)}_{99}$. However, let's take another point of view, which is educated by the equal-mass cases \cite{Pogel:2022ken,Pogel:2022vat,Pogel:2022yat}. The $\varepsilon$-factorised Gauss-Manin connection matrix, denoted as ${\bf A}^{\rm eq}(y)$, enjoys the beautiful self-duality, see eq. (7.2) of \cite{Pogel:2022yat}, which indicates ${\bf A}^{\rm eq}_{2,3} = {\bf A}^{\rm eq}_{3,4}$ and in the context of K3 geometry thereof, and we have ${\bf A}^{\rm eq}_{2,3} = {\bf A}^{\rm eq}_{3,4}=1\cdot {\rm d}\tau$, where $\tau$ is the only moduli in the equal-mass configuration. Here, the unit is simply the trivial $Y$-invariant for the involved K3 surface. In the generic unequal-mass configuration, we have four moduli (or independent kinematic variables), $\{\tau_1, \tau_2, \tau_3, \tau_4\}$. Denote the $\varepsilon$-factorized Gauss-Manin in the current case as ${\bf A}\big(y\big)$, then one of the natural consequences, similar to the equal-mass situation, is the following
\begin{equation}\label{eq:SF}
	{\bf A}_{5,6} =  {\rm d} \tau_1,\quad {\bf A}_{5,7} = {\rm d} \tau_2,\quad {\bf A}_{5,8} = {\rm d} \tau_3,\quad {\bf A}_{5,9} = {\rm d} \tau_4. 
\end{equation}
Now with the $B$-ordered rotation matrix at hand, one can easily check that 
\begin{equation}
	\label{eq:GM1}
	\begin{aligned}
		{\bf A}_{5,5+j} &= \sum_{i=1}^4\frac{R^{(-2)}_{(i+5)(j+5)}}{\psi_0}{\rm d}\log(y_i),\quad j \in \{1, 2, 3, 4\}\,.
	\end{aligned}
\end{equation}
Comparison between \eqref{eq:SF} and \eqref{eq:GM1} leads us to conclude that
\begin{equation}
	\label{eq:SDconsequences}
	\begin{aligned}
		R^{(-2)}_{(i+5)(j+5)} &= y_i\frac{\partial \tau_j}{\partial y_i}\psi_0,\quad i, j \in \{1, 2, 3, 4\}\,,
	\end{aligned}
\end{equation}
where the moduli are defined similarly as the equal-mass case via ratios between single-logarithmic periods $\psi_{1, i}$ and the holomorphic one $\psi_0$:
\begin{equation}\label{eq:moduli}
	\tau_i = \frac{\psi_{1,i}}{\psi_0}\,,\quad  i \in \{1, 2, 3, 4\}\,,
\end{equation}
with $\psi_{1, i}$'s given by \eqref{def_Frobenius}. Then we have 
\begin{equation}
	\label{eq:RFi}
	\begin{aligned}
		64\psi_0 R^{(-2)}_{F(5+j)} =  -\sum_{i=1}^4\left( \partial_i\,+8 \right)\,\psi_0\,R_{(5+i) (5+j)}^{(-2)} = - \sum_{i=1}^4\left(\partial_i\, +8 \right)\,\psi_0^2\,y_i\frac{\partial \tau_j}{\partial y_i},\,\,\, j \in \{1, 2, 3, 4\}. 
	\end{aligned}
\end{equation} 
All the above results pass constraints. 

Now, we are only left with $R_{FF}^{(-2)}$ and $R_{FE}^{(-2)}$. 
We need to solve $R_{FF}^{(-2)}$ first, then solve $R_{FE}^{(-2)}$. They can be systematically solved by series expansions, and read
\begin{equation}
	\begin{aligned}
		R^{(-2)}_{FF} &= 1-4 \big[y_1+y_2+y_3+y_4\big]+\frac{1}{3} \Big[27 \big(y_1^2 +{\rm perm.}\big)+104\big( y_1y_2+{\rm perm.}\big)\Big] + \mathcal{O}(y^3)\,,\\
		R^{(-2)}_{FE} &= \frac{y_3-y_4}{8}\left[3-\left(29y_1 + 22y_2 + 9 (y_3 + y_4)\right) + \mathcal{O}(y^3)\right]\,.
	\end{aligned}
\end{equation}
The asymmetry in $R^{(-2)}_{FE}$ is due to the chosen prefactor in $J_{14}$. 

It is simpler to solve $R^{(-1)}_2$ and $R^{(0)}_2$. On the one hand, there are fewer entangled unknown functions; on the other hand, their constraints can be reduced to inhomogeneous first-order differential equations depending on known functions in $R^{(-2)}_2$. 

To summarise, we give a snapshot of selected elements\footnote{They can be expressed in a closed form using \eqref{def_Frobenius}, which we ignore here. } in the $\eps$-factorised connection matrix, ${\bf A}(y)$, expanded around the maximally unipotent monodromy (MUM) point:
\begin{equation}
	\begin{aligned}
		{\bf A}_{5,5} &= \sum_{i=1}^4 \Big[-6 +22(y_1 + y_2 + y_3 + y_4) - 16 y_i  + \mathcal{O}(y^{-2})\Big]{\rm d} y_i \,,\\
		{\bf A}_{5,6} &= {\rm d}\tau_1\,\\
		&=   \Big[\frac{1}{y_1}+2(y_2+y_3+y_4)  + \mathcal{O}(y^{-2})\Big]{\rm d} y_1\\
		& + \Big[-2 +2(y_1 + y_2 + 4y_3 + 4y_4)  + \mathcal{O}(y^{-2})\Big]{\rm d} y_2\\
		& + \Big[-2 +2(y_1 + 4y_2 + y_3 + 4y_4)  + \mathcal{O}(y^{-2})\Big]{\rm d} y_3\\
		& + \Big[-2 +2(y_1 + 4y_2 + 4y_3 + y_4)  + \mathcal{O}(y^{-2})\Big]{\rm d} y_4\,,\\
		&\vdots 
	\end{aligned}
\end{equation}

\section{Conclusion}
The recently proposed algorithm towards $\eps$-factorising differential equations of Feynman integrals is applicable in a very wide range. To our best knowledge, it is rather general and passes all examples involving geometries. It puts strong evidence on the existence of $\eps$-factorised basis and gives an efficient way to find it. It is very interesting to investigate the deep connection with Hodge theory, to combine it with (semi-)numerical approaches and to apply it to challenging theoretical predictions.

\subsection*{Acknowledgements}

This work has been supported by the Research Unit ``Modern Foundations of Scattering Amplitudes'' (FOR 5582)
funded by the German Research Foundation (DFG).
X.W. is supported by the University Development Fund of The Chinese University of Hong Kong, Shenzhen, under the Grant No. UDF01003912, and by the National Natural Science Foundation of China with Grant No. 12535006.
This research has received funding from the European Research Council (ERC) under the European Union’s Horizon 2022
Research and Innovation Program (ERC Advanced Grant No.~101097780, EFT4jets and ERC Consolidator Grant No.~101043686 LoCoMotive). 
Views and opinions expressed
are however those of the authors only and do not necessarily reflect those of the European Union or the European
Research Council Executive Agency. Neither the European Union nor the granting authority can be held responsible for
them.

%

{\footnotesize
\bibliography{radcor25}
\bibliographystyle{h-physrev5}
}

\end{document}